\documentclass[reprint,showpacs,amsmath,amssymb,aps,prl]{revtex4-1}
\usepackage{graphicx}
\pdfoutput=1
\usepackage{dcolumn}
\usepackage{bm}
\usepackage{hyperref}
\usepackage[mathlines]{lineno}

\begin{document}

\title{Chiral Surface Modes in Three-Dimensional Topological Insulators}
\author{Kiminori Hattori and Hiroaki Okamoto}
\affiliation{Graduate School of Engineering Science, Osaka University, Toyonaka, Osaka 560-8531, Japan}

\begin{abstract}
Where chiral modes should appear is an essential question for the quantum anomalous Hall (QAH) effect in three-dimensional topological insulators (3DTIs).
In this Letter, we show that in a slab of ferromagnetic 3DTI subjected to a uniform exchange field normal to its top and bottom surfaces, the QAH effect creates a single chiral surface mode delocalized on the side faces.
In a nonmagnetic 3DTI, analogously, delocalized helical modes consisting of a pair of oppositely propagating chiral surface modes are produced by the quantum spin Hall effect.
\end{abstract}

\maketitle

The quantum anomalous Hall (QAH) state is an exotic quantum Hall (QH) state without Landau levels.
The QAH effect is theoretically predicted \cite{ref:1,ref:2,ref:3} and experimentally observed \cite{ref:4,ref:5,ref:6} in topological insulators (TIs), which host two-dimensional (2D) Dirac fermions on the surface due to the nontrivial bulk topology.\cite{ref:7,ref:8}
It is well known that in a conventional QH system, 1D chiral edge modes are created in the Landau gap, reflecting the bulk-boundary correspondence.
Analogous chiral modes attributable to the QAH effect attract a great deal of interest in terms of basic physics and potential applications.
Particularly, electron transport via the chiral mode, where backscattering is completely forbidden, is essentially immune to both magnetic and nonmagnetic impurities.
This is a salient feature crucial to low-dissipation electronics.

Regarding the QAH effect in 3DTIs, however, a fundamental property remains unclarified.
The chiral edge state is localized at the circumference of a finite QH system, whereas there are no edges on the closed surface covering a 3DTI.
Where chiral modes should appear is an essential question for the latter.
In a nonmagnetic 3DTI, massless Dirac surface states are formed in the bulk insulating gap.
Incorporating ferromagnetism into the system by magnetic doping results in Dirac fermions being exchange coupled to magnetic moments.
In the presence of an exchange field normal to the surface, the surface spectrum opens a mass gap, yielding the half-quantized Hall conductivity ${e^2}/2h$.\cite{ref:8}
A uniform magnetization generally separates the closed surface of a 3DTI into massive and massless domains.
The previous theoretical study for a 3DTI suggests that the QAH effect produces chiral edge modes that are localized at the interfaces between massive and massless domains.\cite{ref:9}
However, the model assumed in the previous study is semi-infinite in the direction of magnetization. This differs from an experimental geometry which is finite in the direction of magnetization.

The aim of this Letter is to provide the definite answer to the basic question posed above by considering an actual configuration.
We analyze a ferromagnetic 3DTI subjected to a uniform exchange field with realistic parameters in addition to its 2D equivalent as a minimal model.
In a massless domain sandwiched between two massive domains, there exist a single chiral mode and nonchiral quantum well (QW) modes.
The chiral mode is not localized at the mass boundary but essentially delocalized on the 2D plane.
We refer to this as a chiral surface mode to distinguish it from conventional chiral edge modes.
Furthermore, we explore the quantum spin Hall (QSH) effect in a thin slab of nonmagnetic 3DTI, and demonstrate that this effect generates delocalized helical modes constituted of a pair of oppositely propagating chiral surface modes.

We begin by considering Dirac surface states of a semi-infinite 3DTI, which are modeled by the 2D Hamiltonian \cite{ref:7,ref:8}
\begin{equation}
\label{eq:1}
H = \gamma {(\bm{\sigma} \times \mathbf{k})}_z + {m_z}{\sigma_z} ,
\end{equation}
in momentum space, where ${\sigma_\mu}$ (${\mu=x,y,z}$) denotes the Pauli matrix in spin space and ${\gamma/\hbar}$ corresponds to the velocity of surface electrons.
The out-of-plane exchange field ${m_z}$ creates a mass gap of size ${2\left|{m_z}\right|}$ in the Dirac dispersion.
A possible in-plane field ${{\mathbf{m}}_{||}}=({m_x},{m_y},0)$ is ignored for simplicity since it merely shifts the Dirac point by ${\Delta\mathbf{k}=({\mathbf{m}}_{||} \times {\mathbf{z}})/\gamma}$.

\begin{figure}
\centering
\includegraphics{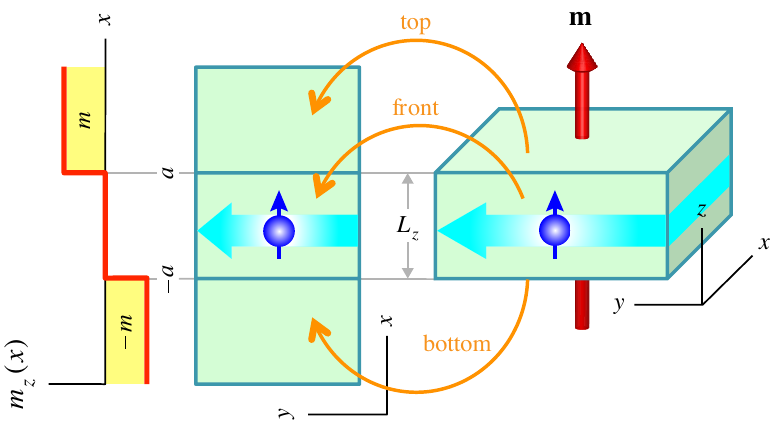}
\caption{(Color online) A slab of ferromagnetic TI subjected to a uniform exchange field ${\mathbf{m}=(0,0,m)}$ and its surface mapped onto the $xy$ plane.
In the 2D model, the out-of-plane field ${m_z}(x)$ is nonuniform.
The massless central region $\left|x\right|<a$ is sandwiched between two massive regions $\left|x\right|>a$ to form a QW structure of width $2a$.
In the two outer regions $x>a$ and $x<-a$, the induced mass $\pm m$ changes sign.
In both models, chiral surface modes propagate in the $y$ direction.}
\label{fig:1}
\end{figure}

As an illustrative example, we suppose a slab of magnetic 3DTI as shown in Fig. \ref{fig:1}.
The TI slab is enclosed by massive top and bottom surface states and massless side surface states.
In this situation, mapping these surfaces onto the $xy$ plane constructs a minimal 2D model, in which $m_z$ in Eq. (\ref{eq:1}) is expressed as ${m_z}(x)=m\operatorname{sgn}(x)\theta(\left|x\right|-a)$.
Note that this procedure is equivalent to unfolding the TI surface.
The gapless central region $\left|x\right|<a$ is sandwiched between the two gapped outer regions $\left|x\right|>a$.
This feature illuminates a QW problem underlying the QAH effect in 3D.

To elucidate microscopic details, we derive the retarded Green's function $G(x,x';{k_y},E)$ at a momentum $k_y$ and an energy $E$ in terms of scattering wave functions following the McMillan method.\cite{ref:10,ref:11,ref:12}
See, Supplemental Material \cite{ref:13} for details of the derivation. 
The local density of states $N(x,E)=\frac{1}{{2\pi}}\int_{-\infty}^\infty {d{k_y}A(x,{k_y},E)}$ and the integrated spectral function $A({k_y},E)=\int_{-b}^{b} {dxA(x,{k_y},E)}$ are calculated from the local spectral function $A(x,{k_y},E)=-\frac{1}{\pi}\operatorname{Im}\operatorname{Tr}G(x,x;{k_y},E)$ in a given sampling section $\left|x\right|<b$.
Similarly, the spin-resolved spectral function ${A_\mu}(x,{k_y},E)=\operatorname{Tr} {\sigma_\mu }A(x,{k_y},E)$ leads to ${N_\mu}(x,E)$ and ${A_\mu}({k_y},E)$.
Spin polarization is characterized by ${P_\mu}(x,E)={N_\mu}(x,E)/N(x,E)$ in real space and ${P_\mu}({k_y},E)={A_\mu}({k_y},E)/A({k_y},E)$ in momentum space.

\begin{figure}
\centering
\includegraphics{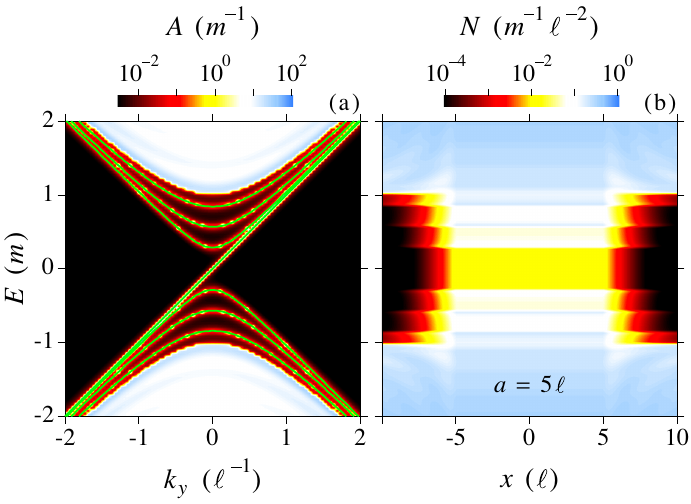}
\caption{(Color online) (a) Spectral function $A({k_y},E)$ and (b) local density of states $N(x,E)$ calculated for the minimal 2D model of $a=5\ell$ and $b=10\ell$.
Solid green lines in (a) represent the QW eigenvalues.}
\label{fig:2}
\end{figure}

Figure \ref{fig:2} shows the numerical results for $m>0$.
In the mass gap $\left|E\right|<\left|m\right|$, the QW effect produces electron and hole subbands.
As expected, the subband dispersion varies appreciably with the QW width $2a$.
A stronger confinement in a narrow well suppresses subband formation in the gap.\cite{ref:13}
Besides these normal QW modes, a single chiral mode traversing the mass gap emerges independently of $a$.
The chiral state, as well as the QW states, spatially extends inside the QW.
Furthermore, it is observed that ${P_x}=1$ and ${P_y}={P_z}=0$ for the chiral mode.\cite{ref:13}

The numerical eigenvalues for bound states in the QW exactly coincide with the spectral peaks shown in Fig. \ref{fig:2}.
Analytically, the QW eigenvalues are found to be $E_n^{(\pm)}=\pm\gamma\sqrt{{k_y}^2+(n\pi/2a)^2}$ ($n=1,2,3,\cdots$) in the $\left|m\right| \to \infty$ limit.
This amounts to the QW gap of size ${E_{{\text{QW}}}}=\pi\gamma/a$ for a sufficiently strong confinement.
For example, ${E_{{\text{QW}}}} \approx 1{\text{meV}}$ at $2a \approx 1\mu{\text{m}}$ with the assumption of $\gamma \approx 0.2{\text{eVnm}}$ (see below).
The finite gap persisting even in such a wide QW reasonably accounts for the scale-invariant QAH effect in 3DTIs observed experimentally.\cite{ref:5}

The chiral mode is also a QW eigenstate.
The relevant eigenvalue problem can be more generically analyzed for an arbitrary mass distribution ${m_z}(x)$.
It is well known that the eigenvalue equation $H(x)\left|{\phi_r}(x)\right\rangle={E_r}\left|{\phi_r}(x)\right\rangle$ possesses a Jackiw-Rebbi solution consisting of ${E_r}=r\gamma {k_y}$ and $\left|{\phi_r}(x)\right\rangle={e^{-\frac{r}{\gamma}\int_0^x {dy{m_z}(y)}}}\left|r\right\rangle$, where $H(x)={e^{-i{k_y}y}}H(x,y){e^{i{k_y}y}}$ and $\left|r\right\rangle=\frac{1}{{\sqrt2}}{(1,r)^t}$ with $r =\pm 1$.\cite{ref:14}
For ${m_z}(x)$ in the 2D model, only $r=\operatorname{sgn}(m)$ gives a nondivergent physical solution, which is explicitly expressed as $\left|{\phi(x)}\right\rangle={e^{-(\left|x\right|-a)\theta(\left|x\right|-a)/\ell}}\left|{\operatorname{sgn}(m)}\right\rangle$, where $\ell=\gamma/\left|m\right|$ is the characteristic length scale of the QW confinement.
Thus, the chiral-state wave function is essentially uniform in the massless central region $\left|x\right|<a$, and exponentially decays in the two massive outer regions $\left|x\right|>a$.
The chiral surface mode is a natural extension of the chiral interface mode described by $\left|{\phi}(x)\right\rangle={e^{-\left|x\right|/\ell}}\left|{\operatorname{sgn}(m)}\right\rangle$ in the $a \to 0$ limit.\cite{ref:12}
It is also worth noting that $\left|r\right\rangle$ corresponds to the eigenspinor of ${\sigma_x}$, i.e., the chiral surface mode propagating in the $\pm y$ direction is perfectly spin-polarized in the $\pm x$ direction.
This is a corollary of the spin-velocity locking prescribed by ${\mathbf{v}}=\frac{\gamma}{\hbar}{\mathbf{z}} \times {\bm{\sigma}}$ for a 2D Dirac fermion on the TI surface.

The presence or the absence of chiral modes can be examined from the viewpoint of charge conservation.
In response to electromagnetic fields $\mathbf{E}$ and $\mathbf{B}$, the QAH effect induces local charge ${\rho_{\text{H}}}={\sigma_{xy}}{B_z}$ and current ${\mathbf{j}}_{\text{H}}={\sigma_{xy}}\mathbf{E} \times \mathbf{z}$.
They satisfy a generalized continuity equation ${\partial_t}{\rho_{\text{H}}}+\nabla\cdot {\mathbf{j}}_{\text{H}}=g_{\text{H}}$.
The source term $g_{\text{H}}=\mathbf{E} \cdot (\mathbf{z} \times \nabla\sigma_{xy})$ is nonvanishing if the QAH conductivity $\sigma_{xy}=-\frac{e^2}{2h}\operatorname{sgn}({m_z})$ varies in space.\cite{ref:12}
It is easily found that the induced charge $Q_{\text{H}}$ follows the relation $\frac{d}{dt}{Q_{\text{H}}}=\int_{-\infty }^{\infty} {dx{g_{\text{H}}}}=-\frac{e^2}{h}r{E_y}$, where $r=\frac{1}{2}\{\operatorname{sgn}[{m_z}(\infty)]-\operatorname{sgn}[{m_z}(-\infty)]\}$.
This means that mass inversion such that ${m_z}(\infty){m_z}(-\infty)<0$ always requires a single chiral mode obeying $\frac{d}{dt}{Q_{\text{C}}}=\frac{e^2}{h}r{E_y}$ to conserve the total charge $Q=Q_{\text{H}}+Q_{\text{C}}$.
The analytical and numerical results described above support this argument.

The minimal 2D model is useful in obtaining physical insight into the chiral surface mode confined in a QW.
However, this model is not enough to quantitatively describe the 3D nature of a realistic sample.
In what follows, we deal with such a 3D problem.
For this purpose, we employ the $4 \times 4$ Dirac Hamiltonian representing a ferromagnetic 3DTI, which is given in momentum space by
\begin{equation}
\label{eq:2}
H = D({\mathbf{k}})+{\mathbf{A}}({\mathbf{k}}) \cdot {\tau_x}{\bm{\sigma}}+B({\mathbf{k}}){\tau_z}+m{\sigma_z} ,
\end{equation}
where $D={D_1}{k_z}^2+{D_2}{k^2}$, ${\mathbf{A}}=({A_2}{k_x},{A_2}{k_y},{A_1}{k_z})$, $B={B_0}-{B_1}{k_z}^2-{B_2}{k^2}$, ${k^2}={k_x}^2+{k_y}^2$, and $\tau_\mu$ denotes the Pauli matrix in orbital space.
The parameters are estimated in the literature for the Bi$_2$Se$_3$ family of TI materials.\cite{ref:7}

For numerical calculation, Eq. (\ref{eq:2}) is discretized on a cubic lattice with lattice spacing $a$.
The lattice Hamiltonian $H=\sum\nolimits_j {H_j}$ is decomposed into ${H_j}=\left| j \right\rangle H_0 \left\langle j \right|+\left| j \right\rangle V \left\langle j + 1 \right|+\left| j + 1 \right\rangle V^\dag \left\langle j \right|$, where $H_0$ denotes the Hamiltonian matrix for a single isolated slice parallel to the $yz$ plane, and $V$ is the hopping matrix connecting the two adjacent slices along $x=ja$.
The periodic boundary condition along $y$ is imposed to remove side surfaces normal to $y$.
To eliminate coupling between side surfaces normal to $x$, we assume the semi-infinite region $x \in (0,\infty)$.
The intraslice Green's function ${G_{jj}}=\left\langle j \right| (E - H)^{-1} \left| j \right\rangle$ can be computed recursively as follows: ${G_{jj}}={({g_j}^{-1}-Vg{V^\dag})^{-1}}$ and ${g_j}={({g_0}^{-1}-{V^\dag}{g_{j-1}}V)^{-1}}$ with ${g_0}={(E-{H_0})^{-1}}$.
The surface Green's function $g={G_{11}}$ of the semi-infinite region is derived by numerically solving the quadratic matrix equation $g={g_0}+{g_0}Vg{V^\dag}g$.\cite{ref:15,ref:16,ref:17}
The present recursive procedure is similar to that devised previously \cite{ref:18} but is more efficient in calculating the local Green's function $G_{jj}$.
The local spectral function is given by $A({\mathbf{r}},{k_y},E)=-\frac{1}{\pi {a^2}}\operatorname{Im}\operatorname{Tr}{G_{jj,ll}}({k_y},E)$, where ${\mathbf{r}}=(x,z)=a(j,l)$.
From $A({\mathbf{r}},{k_y},E)$ in a given sampling section $x \in (0,d)$, we obtain the local density of states $N({\mathbf{r}},E)=\frac{1}{{2\pi }}\int_{-\pi/a}^{\pi/a} {d{k_y}A({\mathbf{r}},{k_y},E)} $ and the integrated spectral function $A({k_y},E)=\int_0^d {dx\int_0^{{L_z}}{dzA({\mathbf{r}},{k_y},E)}}$.
We also examine the local current density defined by ${J_y}({\mathbf{r}},E)=\int_{-\pi/a}^{\pi/a}{d{k_y}\operatorname{Tr}\hbar{v_y}A({\mathbf{r}},{k_y},E)}$ with $\hbar{v_y}=\partial H/\partial {k_y}$ in addition to the spin polarization introduced above.
Note that for a chiral mode, the spatial integration ${G_y}=\int_0^d {dx\int_0^{{L_z}} {dz{J_y}({\mathbf{r}},E)}}$ corresponds with two-terminal conductance in units of ${e^2}/h$ in the $d \to \infty$ limit.
An analogous recursive algorism applies to the opposite semi-infinite region $x \in (-\infty,0)$.
The numerical results shown below are obtained with $a=1{\text{{\AA}}}$ and realistic parameters for Bi$_2$Se$_3$.

\begin{figure}
\centering
\includegraphics{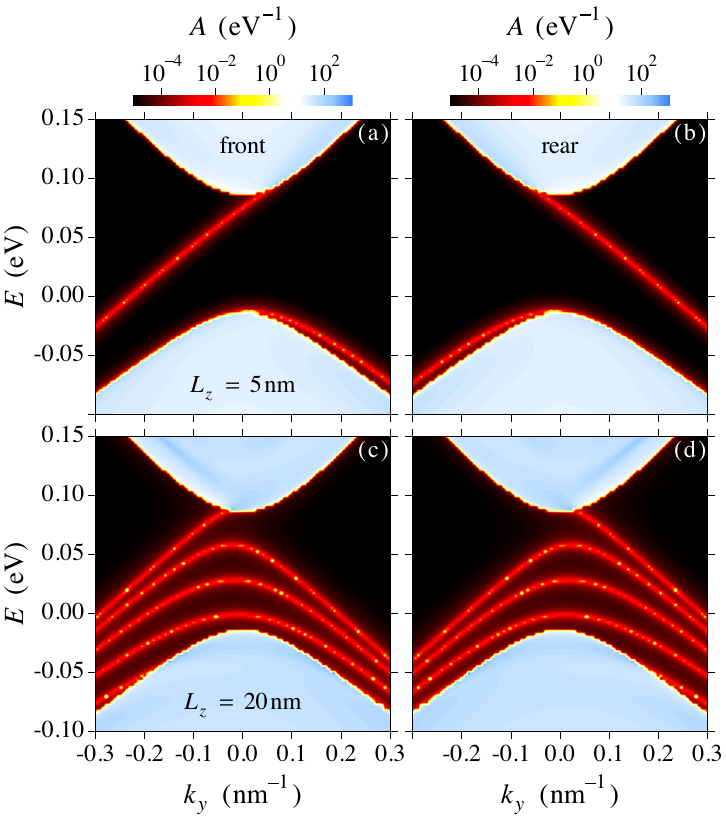}
\caption{(Color online) Spectral function $A({k_y},E)$ calculated for a semi-infinite ferromagnetic TI with $m=50{\text{meV}}$.
The parameters are ${L_z}=5{\text{nm}}$ in [(a) and (b)] and 20nm in [(c) and (d)].
[(a) and (c)] show the results for the front sampling section, and [(b) and (d)] for the rear sampling section.
The sampling length is set at $d=10{\text{nm}}$.}
\label{fig:3}
\end{figure}

\begin{figure}
\centering
\includegraphics{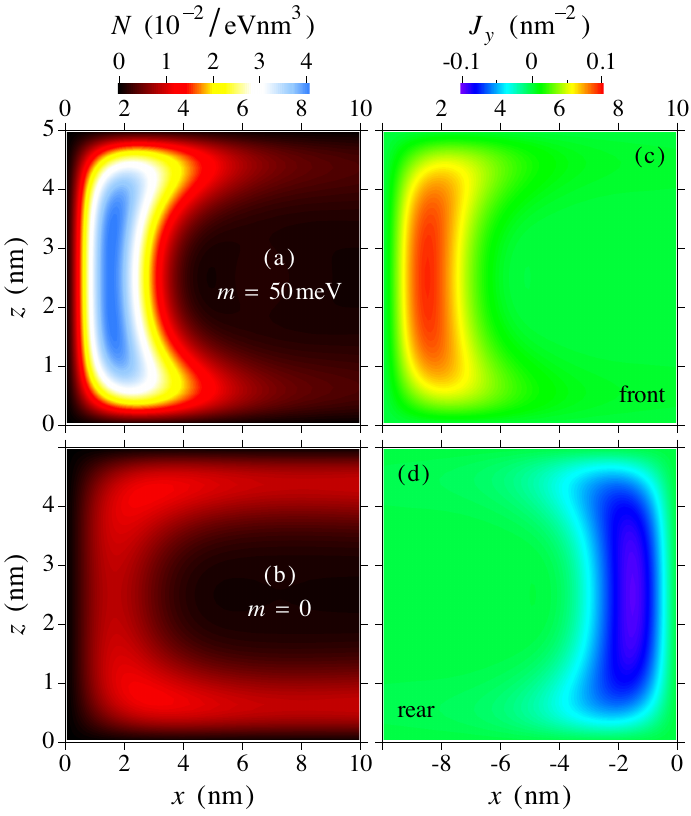}
\caption{(Color online) Local density of states $N(x,z)$ and local current density ${J_y}(x,z)$ at $E=50{\text{meV}}$ calculated for a semi-infinite TI.
The parameters are ${L_z}=5{\text{nm}}$ and $d=10{\text{nm}}$.
The left two graphs compare $N(x,z)$ for (a) $m=50{\text{meV}}$ and (b) $m=0$ in the front sampling section.
The right two graphs show ${J_y}(x,z)$ for $m=50{\text{meV}}$ on (c) front and (d) rear sides.}
\label{fig:4}
\end{figure}

Figure \ref{fig:3} displays $A({k_y},E)$ calculated for $m=50{\text{meV}}$ and ${L_z} \geqslant 5{\text{nm}}$.
In this thickness range, intersurface coupling is negligibly weak relative to the exchange interaction so that the system is in the QAH phase.\cite{ref:13}
As seen in the figure, spectral peaks are comprised of a single chiral mode and multiple QW modes in the mass gap.
The observed QW modes consist only of hole subbands.
The lack of electron subbands is due to electron-hole asymmetry induced by the $D$ term in Eq. (\ref{eq:2}).
The QW modes are eliminated by sufficiently reducing $L_z$, leaving only a chiral mode across the gap.
The chiral mode propagates in opposite directions on front [$x \in (0,d)$ in $(0,\infty)$] and rear [$x \in (-d,0)$ in $(-\infty,0)$] sides, implying its unidirectional circulation in a finite system.
Figure \ref{fig:4} compares $N({\mathbf{r}},E)$ at $E=50{\text{meV}}$ for $m=50{\text{meV}}$ to that for $m=0$.
Metallic surface states continuously enclose the inner bulk region for $m=0$, whereas an extended state sticks to the side surface in the QAH state.
The latter is characteristic of a chiral surface mode.
The local currents ${J_y}({\mathbf{r}},E)$ are also shown in this figure.
The spatial distribution of ${J_y}$ is very similar to that of $N$ inside the QW gap.
As expected, the signs of ${J_y}$ are opposite on front and rear sides.
Moreover, it is verified that $\left| {G_y} \right| \simeq 1$.
These results corroborate that there exists a single chiral surface mode.
On the other hand, ${J_y}$ tends to spatially oscillate for out-gap states.\cite{ref:13}	

These observations can be understand quantitatively in terms of the effective 2D Hamiltonian deduced from projecting Eq. (\ref{eq:2}) onto the subspace of surface states.\cite{ref:7,ref:19,ref:20,ref:21}
The effective Hamiltonian is given by ${H_{xy}}=\gamma {(\bm{\sigma} \times \mathbf{k})}_z {s_z}+m{\sigma_z}$ for decoupled top and bottom surface states, for which ${s_z}=\pm 1$, respectively.
The parameter is evaluated to be $\gamma={A_2}\sqrt{1-{I_1}^2}=0.41{\text{eVnm}}$ and ${I_1}={D_1}/{B_1}={\text{0.13}}$, yielding an estimate for the chiral-state decay length at $\ell=\gamma/m=8.1{\text{nm}}$.
This explains $N({\mathbf{r}},E)$ near $z=0$ or ${L_z}$, which forms a tail away from the side surface of an extent on the order of $\ell/2$.
The anisotropic Hamiltonian expressed as ${H_{yz}}=({\gamma_1}{\sigma_y}{k_z}-{\gamma_2}{\sigma_z}{k_y}){s_z}+{I_2}m{\sigma_z}$ is derived for front (${s_z}=-1$) and rear (${s_z}=+1$) side surface states.
The parameters are given by ${\gamma_1}={A_1}\sqrt{1-{I_2}^2}=0.21{\text{eVnm}}$, ${\gamma_2}={A_2}\sqrt{1-{I_2}^2}=0.38{\text{eVnm}}$ and ${I_2}={D_2}/{B_2}=0.35$.
Following the minimal 2D model, the QW level spacing amounts to $\Delta E=\pi {\gamma_1}/{L_z}$.
This corresponds with $\Delta E \approx 30{\text{meV}}$ observed for ${L_z}=20{\text{nm}}$.
The exchange field $\mathbf{m}$ is parallel to side surfaces.
The in-plane field leads to opposite momentum shifts by $\Delta {k_y}={I_2}m{s_z}/{\gamma_2}$ for ${s_z}=\pm 1$, being consistent with the observation.
Moreover, the spin-velocity locking ${v_y}=-{\gamma_2}{s_z}{\sigma_z}/\hbar$ on side surfaces implies that the chiral mode is spin polarized in the $+z$ direction on both front and rear sides.
This is confirmed in the numerical calculation.\cite{ref:13}

Finally, we address the QSH effect in a thin slab of nonmagnetic 3DTI.\cite{ref:19,ref:20,ref:21}
The relevant 2D Hamiltonian is formulated as ${H_{xy}}=\gamma {(\bm{\sigma} \times \mathbf{k})}_z {s_z}+t{s_x}$, where $s_\mu$ denotes the Pauli matrix in orbital space spanned by top and bottom surface states, and $t=\Delta -B{k^2}$ represents tunneling coupling between these two states.
The intersurface mixing opens a gap of size $2\left|\Delta\right|$ in the Dirac dispersion.
The $H_{xy}$ given above is block-diagonalized by orthogonal transformation, leading to the $2 \times 2$ Hamiltonians ${H_\pm}=\gamma {(\bm{\sigma} \times \mathbf{k})}_z \pm t{\sigma_z}$ that describe two independent subsystems with Chern numbers ${C_\pm } = \mp \frac{1}{2}(\operatorname{sgn}\Delta+\operatorname{sgn}B)$.
In the parameter range where $\Delta B>0$, the entire system is in the QSH phase that hosts a helical pair of two chiral edge modes running in opposite directions.
Although the 2D model captures these general features of the QSH effect, it neglects side surface states in a TI slab and hence cannot describe the associated 3D characteristics.
In particular, it is unclear in this model whether there exist helical surface modes analogous to chiral surface modes.

\begin{figure}
\centering
\includegraphics{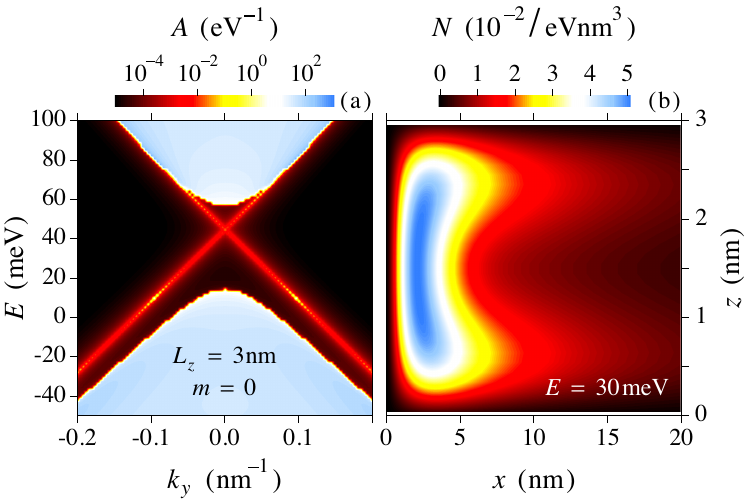}
\caption{(Color online) (a) Spectral function $A({k_y},E)$ and (b) local density of states $N(x,z)$ at $E=30{\text{meV}}$ calculated for a semi-infinite nonmagnetic TI of ${L_z}=3{\text{nm}}$.
The sampling length is set at $d=20{\text{nm}}$.
The spectral functions are identical for the front and rear sampling sections.}
\label{fig:5}
\end{figure}

Figure \ref{fig:5} summarizes the numerical results for a nonmagnetic 3DTI of thickness ${L_z}=3{\text{nm}}$.
At this thickness, $\Delta =-20{\text{meV}}$ and $B=-0.2{\text{eV}}{{\text{nm}}^2}$ so that the system is in the QSH phase.\cite{ref:14,ref:21}
The QSH state is manifested by helical modes across the hybridization gap.
Similar to chiral surface states, the helical states are delocalized on the side surface.
The 2D model predicts helical edge states of decay length $\sim26{\text{nm}}$.\cite{ref:21}
This does not contradict with the observed $N({\mathbf{r}},E)$ slowly decaying along $x$.
The spin-velocity locking results in opposite spin polarizations for oppositely propagating chiral modes.
The antiparallel polarization of helical surface modes is confirmed in the numerical calculation, which shows that ${P_x}={P_y}=0$, and $\operatorname{sgn}{P_z}= \pm \operatorname{sgn}{v_y}$ on front and rear sides, respectively.\cite{ref:13}

In summary, we have investigated the QAH and QSH effects in 3DTIs with realistic parameters.
The QAH effect generates a chiral surface mode delocalized in a massless domain sandwiched between two massive domains on the surface of a ferromagnetic 3DTI.
In a nonmagnetic 3DTI, helical surface modes consisting of a pair of oppositely propagating chiral surface modes are created by the QSH effect.

The Jackiw-Rebbi model indicates that the chiral surface mode remains delocalized in an arbitrarily thick slab of ferromagnetic 3DTI.
The spatial profile of this mode could be probed by the scanning tunneling microscopy or the spatially resolved Kerr rotation technique.
The microscopic details are of practical importance for the QAH experiment and potential applications of the robust QAH conduction, both of which generally require an effective coupling between propagating modes and external leads.

\bibliography{ref}

\pagebreak

\appendix

\section{Supplemental Material for \\ ``Chiral Surface Modes in Three-Dimensional Topological Insulators''}

In this supplemental material, we provide additional explanations for the Green's function and QW eigenvalues in the 2D model, intersurface mixing in a slab of 3DTI, spin polarization of chiral and helical surface modes, and energy dependence of local current.

\section{Green's function}
The retarded Green's function for the 2D model can be formulated in terms of scattering wave functions.
Because of translational invariance along $y$, the surface Hamiltonian is reduced to $H(x)={e^{-i{k_y}y}}H(x,y){e^{i{k_y}y}}$ for a plane wave solution.
In three separate spatial regions $x<-a$, $\left|x\right|<a$ and $x>a$ (labeled with $j=1,2,3$, respectively), the eigenfunctions are given by $\phi_j^{(\pm)}(x)={(c_j^{(\pm)},1)^t}{e^{\pm i{k_j}x}}$, where $c_j^{(\pm)}=\frac{{\gamma ({k_y} \pm i{k_j})}}{{E-{m_j}}}$, $\gamma {k_j}=\sqrt{{E^2}-{\gamma^2}k_y^2-m_j^2}$, $m=-{m_1}={m_3}$ and ${m_2}=0$.
Assembling these local eigenmodes, the scattering wave function for upward incidence from the region 1 is represented as
\begin{equation*}
\phi_U^{(e)}(x) = \begin{cases}
{\phi_1^{(+)}(x)+r_U^{(e)}\phi_1^{(-)}(x)} &{(x \in 1)} \\ {f_U^{(e)}\phi_2^{(+)}(x)+b_U^{(e)}\phi_2^{(-)}(x)} &{(x \in 2)} \\ {t_U^{(e)}\phi_3^{(+)}(x)} &{(x \in 3)} \end{cases} ,
\end{equation*}
in the electron-like region $E>0$.
The reflection and transmission coefficients are determined from the continuity of wave functions at $x=\pm a$ to be $f_U^{(e)}=\frac{{t_{12}^{(e)}}}{{{\eta^{(e)}}}}{e^{i({k_2}-{k_1})a}}$, $b_U^{(e)}=\frac{{r_{23}^{(e)}t_{12}^{(e)}}}{{{\eta^{(e)}}}}{e^{i(3{k_2}-{k_1})a}}$, $r_U^{(e)}=(r_{12}^{(e)}+\frac{{r_{23}^{(e)}t_{12}^{(e)}t_{21}^{(e)}}}{{{\eta^{(e)}}}}{e^{4i{k_2}a}}){e^{-2i{k_1}a}}$ and $t_U^{(e)}=\frac{{t_{12}^{(e)}t_{23}^{(e)}}}{{{\eta^{(e)}}}}{e^{i(2{k_2}-{k_1}-{k_3})a}}$.
Here, the elemental scattering coefficients are given by $r_{ij}^{(e)}=\frac{{c_i^{(s)}-c_j^{(s)}}}{{c_j^{(s)}-c_i^{(-s)}}}$ and $t_{ij}^{(e)}=1+r_{ij}^{(e)}$ with $s=\operatorname{sgn}(j-i)$.
The denominator ${\eta^{(e)}}=1-r_{21}^{(e)}r_{23}^{(e)}{e^{4i{k_2}a}}$ describes resonant multiple reflections in the QW.
In terms of this, the eigenvalue equation for QW bound states is simply expressed as ${\eta^{(e)}}=0$.
An analogous formulation is derived for the scattering wave function $\phi_D^{(e)}$ for downward incidence from the region 3.

The Green's function obeys the equation of motion $[E-H(x)]G(x,x')=\delta(x-x')$.
The solution is found to be
\begin{equation*}
G_{\sigma \sigma'}^{(e)}(x,x')=\frac{{{C^{(e)}}}}{\gamma} \times \begin{cases}
  {\phi_{U\sigma}^{(e)}(x)\phi_{D\sigma'}^{(e)}(x')}&{(x>x')} \\ 
  {\phi_{D\sigma}^{(e)}(x)\phi_{U\sigma'}^{(e)}(x')}&{(x<x')} 
\end{cases},
\end{equation*}
where $\sigma=\uparrow,\downarrow$ denotes the spin index.
The normalization constant ${C^{(e)}}=1/(c_3^{(+)}-c_3^{(-)})t_U^{(e)}$ is derived from the boundary condition for $G(x,x')$ at $x=x'$.
${G^{(h)}}(x,x')$ in the hole-like region $E<0$ is obtained by replacing $\pm{k_j} \to \mp{k_j}$, and accordingly $c_j^{(\pm)} \to c_j^{(\mp)}$.
The present analysis is equally applicable to propagating and evanescent modes under the condition $\operatorname{sgn}(E)\operatorname{Im}{k_j}>0$, which establishes the asymptotic behavior ${\phi_U}(+\infty)={\phi_D}(-\infty)=0$, i.e., $G(\pm\infty,\mp\infty)=0$.

\section{QW eigenvalues}
\begin{figure}
\centering
\includegraphics{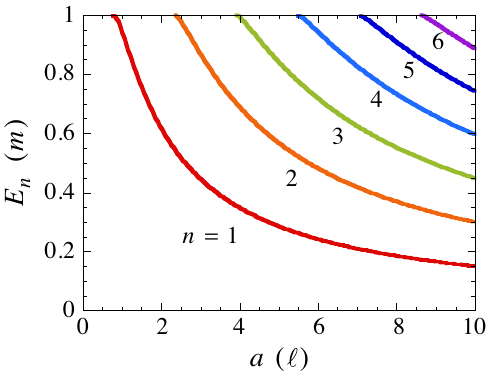}
\caption{QW eigenvalues ${E_n}$ at ${k_y}=0$ calculated as a function of $a$.}
\label{fig:S1}
\end{figure}

The eigenvalue equation for QW bound states is given by ${\eta^{(e)}}={\eta^{(h)}}=0$.
Figure \ref{fig:S1} summarizes positive eigenvalues ${E_n}$ ($n=1,2,3,\cdots$) for electron subbands at ${k_y}=0$ as a function of $a$.
Note that negative eigenvalues $-{E_n}$ are relevant to hole subbands in consequence of electron-hole symmetry.
The QAH conduction via chiral mode is observable in the QW gap $\left|E\right|<{E_1}$.
As $\left|m\right|\to\infty$, $r_{21}^{(e)}r_{23}^{(e)}$ approaches unity so that ${E_n}=\gamma\sqrt{{k_y}^2+{{(n\pi/2a)}^2}}$.

\section{Intersurface mixing}

\begin{figure}
\centering
\includegraphics{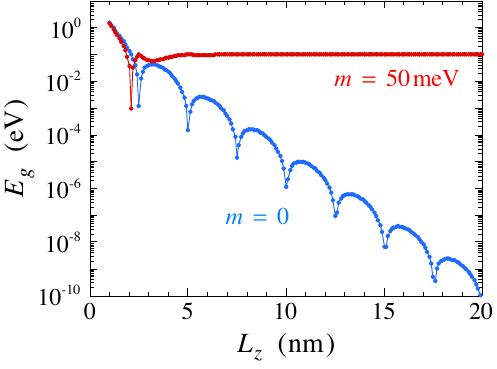}
\caption{Band gap ${E_g}$ of a TI slab as a function of ${L_z}$ for $m=0$ and 50meV.}
\label{fig:S2}
\end{figure}

Figure \ref{fig:S2} displays the band gap ${E_g}$ of a TI slab as a function of ${L_z}$ calculated for $m=0$ and 50meV.
For $m=0$, intersurface mixing creates a finite gap, which exhibits an oscillatory exponential decay with $L_z$ [19-21].
This behavior is approximately described by ${E_g}=2\left|\Delta \right|$ and $\Delta \propto {e^{-\alpha {L_z}}}\operatorname{sin}\beta {L_z}$, where $\alpha = {A_1}/2\sqrt{{B_1}^2-{D_1}^2}$ and $\beta = \sqrt{{B_0}/{B_1}-{\alpha^2}}$.
The oscillation period $\pi /\beta$ is estimated to be 2.5nm for Bi$_2$Se$_3$.
For $m \ne 0$, ${E_g}$ is expressed generally as ${E_g}=2\operatorname{min}\left|{\Delta \pm m}\right|$.
For ${L_z}\geqslant5{\text{nm}}$ and $m=50{\text{meV}}$, $\left|\Delta\right|$ is orders of magnitudes smaller than $m$ and hence ${E_g}\cong2\left|{m}\right|$.
Thus, the QAH criterion $\left|{m}\right|>\left|\Delta\right|$ [2,3,21] is reasonably fulfilled under these conditions.

\section{Spin polarization}
\begin{figure}
\centering
\includegraphics{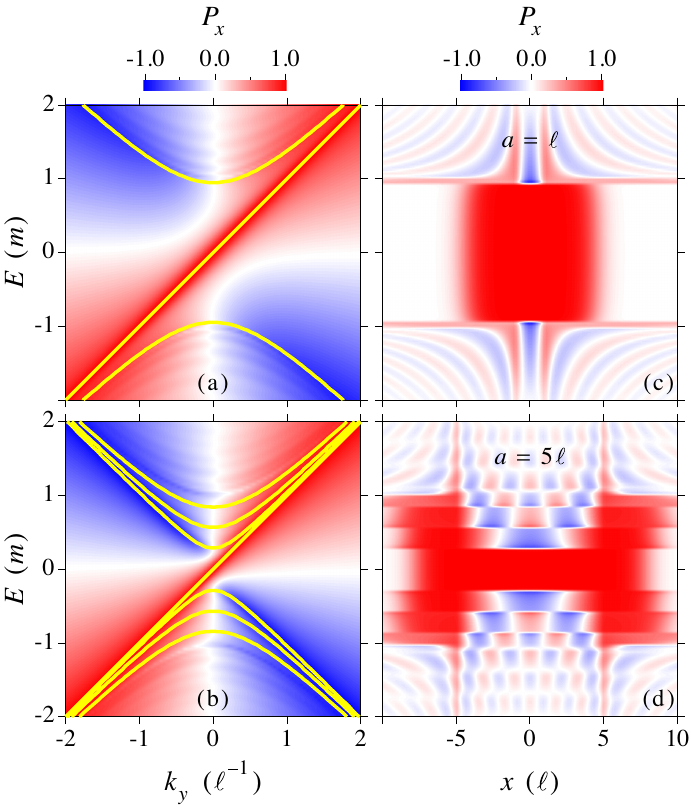}
\caption{Spin polarization [(a) and (b)] ${P_x}({k_y},E)$ and [(c) and (d)] ${P_x}(x,E)$ calculated for the 2D model.
The parameters are $a=\ell$ in [(a) and (c)] and $a=5\ell$ in [(b) and (d)].
The sampling length is set at $b=10\ell$.
Solid yellow lines in [(a) and (b)] represent the QW eigenvalues.}
\label{fig:S3}
\end{figure}

\begin{figure}
\centering
\includegraphics{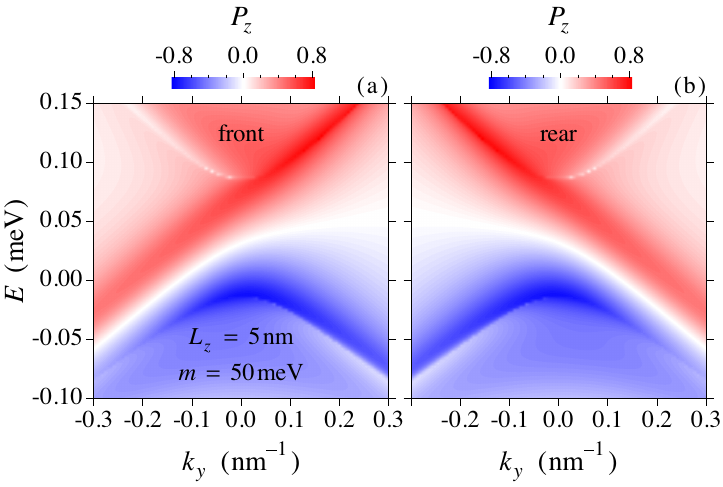}
\caption{Spin polarization ${P_z}({k_y},E)$ calculated for a semi-infinite ferromagnetic TI.
(a) and (b) show the results for the front and rear sampling sections, respectively.
The parameters are ${L_z}=5{\text{nm}}$, $m=50{\text{meV}}$ and $d=10{\text{nm}}$.}
\label{fig:S4}
\end{figure}

\begin{figure}
\centering
\includegraphics{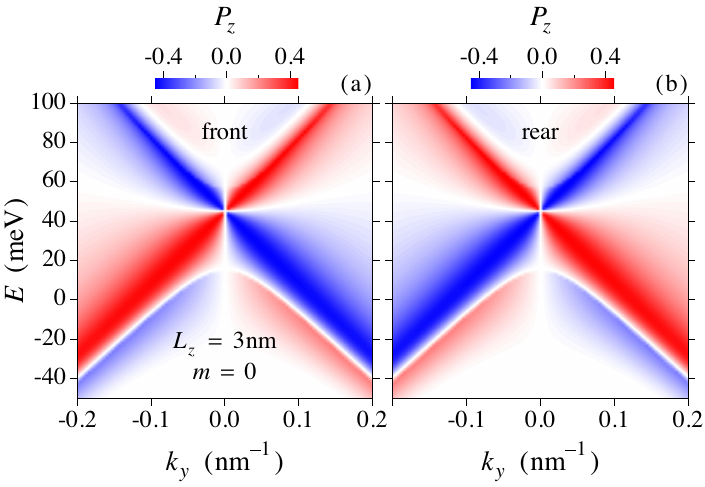}
\caption{Spin polarization ${P_z}({k_y},E)$ calculated for a semi-infinite nonmagnetic TI.
(a) and (b) show the results for the front and rear sampling sections, respectively.
The parameters are ${L_z}=3{\text{nm}}$, $m=0$ and $d=20{\text{nm}}$.}
\label{fig:S5}
\end{figure}

Figure \ref{fig:S3} shows ${P_x}({k_y},E)$ and ${P_x}(x,E)$ for the 2D model assuming $m>0$.
It is clear in the figure that ${P_x}=1$ for a chiral surface mode.
Figure \ref{fig:S4} displays ${P_z}({k_y},E)$ at $m=50{\text{meV}}$ for the 3D model.
This demonstrates that a chiral surface mode possesses spin polarization pointing in the $+z$ direction on both front and rear sides.
For helical surface modes observed at $m=0$, the relations $\operatorname{sgn}{P_z}= \pm \operatorname{sgn}{v_y}$ hold on front and rear sides, respectively, as shown in Fig. \ref{fig:S5}.
All of these observations are basically accounted for by the spin-velocity locking formulated in the text.
For the 3D model, generally $\left|{{P_z}}\right|<1$.
This property may be ascribed to the effective in-plane spin operator ${I_2}{\sigma_z}$ (${I_2}=0.35$) renormalized for side surface states.

\section{Local current}

\begin{figure}
\centering
\includegraphics{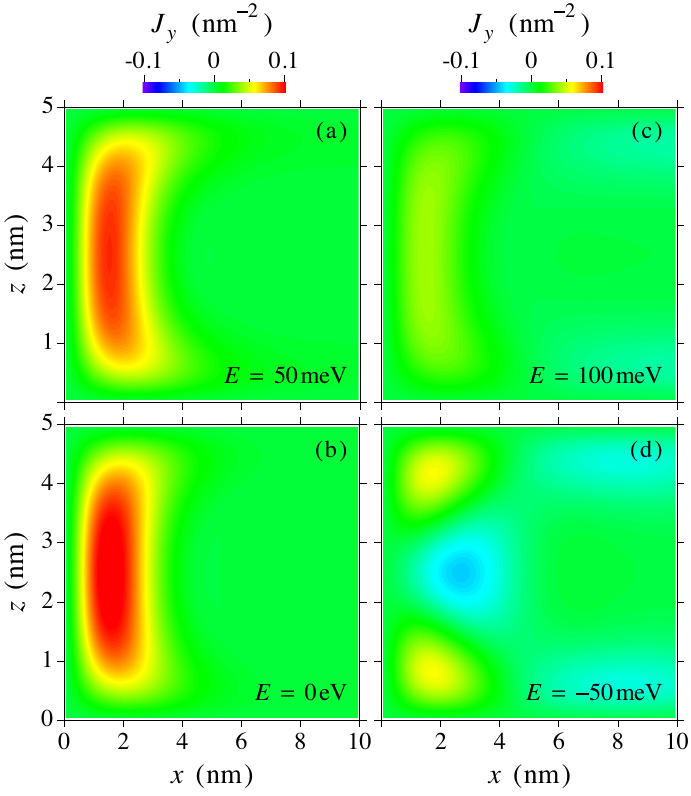}
\caption{Local currents ${J_y}(x,z)$ calculated for a semi-infinite ferromagnetic TI in the front sampling section at (a) $E=50{\text{meV}}$, (b) 0eV, (c) 100meV and (d) -50meV.
The parameters are ${L_z}=5{\text{nm}}$, $m=50{\text{meV}}$ and $d=10{\text{nm}}$.}
\label{fig:S6}
\end{figure}

\begin{figure}
\centering
\includegraphics{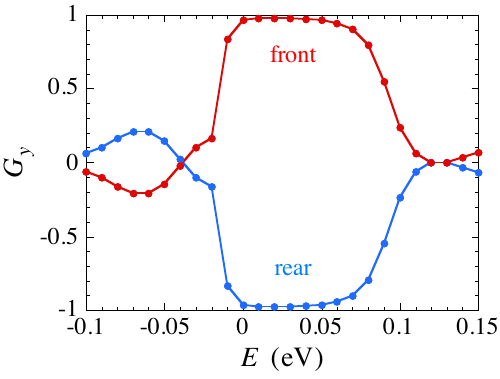}
\caption{Integrated current ${G_y}$ as a function of energy for a semi-infinite ferromagnetic TI.
The parameters are ${L_z}=5{\text{nm}}$, $m=50{\text{meV}}$ and $d=10{\text{nm}}$.}
\label{fig:S7}
\end{figure}

Figure \ref{fig:S6} summarizes local currents ${J_y}({\mathbf{r}},E)$ in the front sampling section of the 3D model at various energies $E$.
Inside the QW gap [(a) and (b)], ${J_y}$ is almost energy independent, and its spatial profile is very similar to that of local density of states $N$.
It is also seen in Fig. \ref{fig:S7} that $\left| {G_y} \right|$ is close to unity.
These features are retained in the gap irrespective of ${L_z}$.
On the other hand, ${J_y}$ tends to spatially oscillate outside the gap, as shown in Fig. \ref{fig:S6} (c) and (d).

\begin{figure}
\centering
\includegraphics{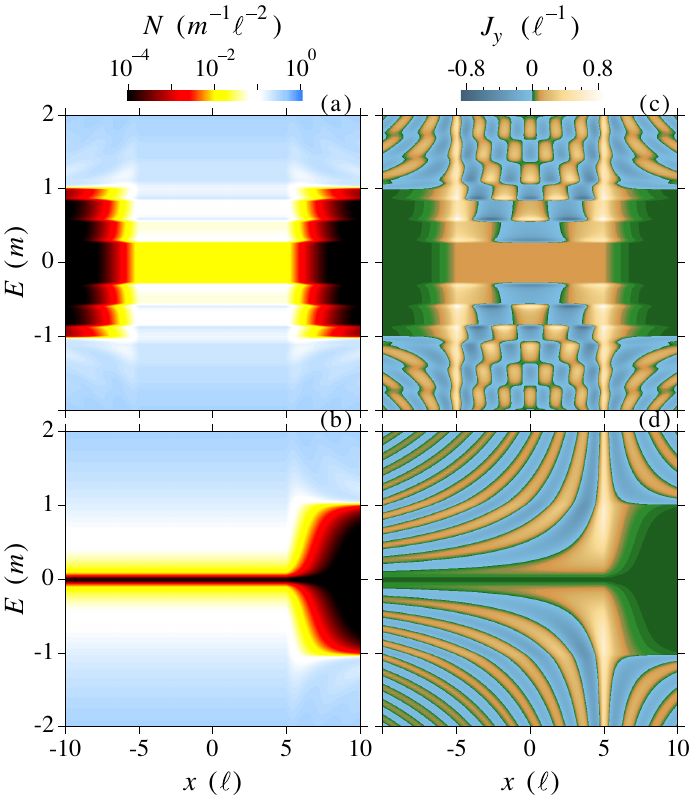}
\caption{[(a) and (b)] Local density of states $N(x,E)$ and [(c) and (d)] local current density ${J_y}(x,E)$ calculated for the 2D model.
The mass distributions are assumed to be ${m_z}(x)=m \operatorname{sgn}(x)\theta(\left|x\right|-a)$ in [(a) and (c)] and ${m_z}(x)=m\theta (x-a)$ in [(b) and (d)], for both of which $a=5\ell$.
Note that in (b), $N$ vanishing at $E=0$ for $x<a$ is characteristic of the density of massless Dirac surface states $\left| E \right|/2\pi {\gamma^2}$.}
\label{fig:S8}
\end{figure}

A similar behavior is observed for the minimal 2D model, as illustrated in Fig. \ref{fig:S8}.
Note that for the 2D model, ${J_y}=2\pi\gamma {N_x}$ reflects the spin density of states.
As shown in (c), the local current $J_y$ carried by the chiral surface mode is spatially uniform inside the QW gap.
The Jackiw-Rebbi solution for the QW ensures that this behavior is size independent and hence observable for an arbitrarily large but finite width $a$.
Outside the gap, on the other hand, there exist QW subbands or conduction and valence bands, for which $J_y$ exhibits an oscillatory pattern.
In the $a \to \infty$ limit, the gap vanishes, and the QW states merge into gapless surface states.
This limit is formally treated by an asymmetric mass distribution such that ${m_z}(x)=m\theta (x-a)$, for which ${m_1}={m_2}=0$ and ${m_3}=m$.
As demonstrated in (d), in this case, the amplitude of $J_y$ is largest at the mass boundary $x=a$ and gradually decays as $x \to -\infty$.
This observation reasonably accounts for ``surface edge states" derived in the previous study assuming an infinitely extended massless region (see, Fig. 4 of Ref. 9).

\end{document}